\documentclass[12pt]{article}
\usepackage{amsfonts,amsmath}
\usepackage[mathscr]{eucal}
\usepackage{amssymb}
\usepackage{amsthm}
\theoremstyle{plain}

\usepackage{epsfig,epsf}
\usepackage{graphicx, subfigure}
\usepackage{graphicx}

\newcommand{\be}{\begin{eqnarray}}
\newcommand{\ee}{\end{eqnarray}}
\newcommand{\nn}{\nonumber \\}

\newcommand{\p}[1]{(\ref{#1})}
\newcommand{\vecg}[1]{\mbox{\boldmath $#1$}}

\begin{document}

\begin{titlepage}

\vfill
\vfill

\begin{center}
\baselineskip=16pt {\Large  Witten index in ${\cal N} = 1$ and ${\cal N} = 2$
SYMCS theories with matter. }
\vskip 0.3cm {\large {\sl }}
\vskip 10.mm {\bf 
A.V. Smilga}
 \\
\vskip 1cm

SUBATECH, Universit\'e de Nantes, \\
4 rue Alfred Kastler, BP 20722, Nantes 44307, France
\footnote{On leave of absence from ITEP, Moscow, Russia}\\
E-mail:  smilga@subatech.in2p3.fr

\end{center}

\vspace{0.2cm} \vskip 0.6truecm \nopagebreak

\begin{abstract}
\noindent
We calculate  the Witten index for 3d  
supersymmetric Yang-Mills-Chern-Simons theories with matter. For ${\cal N} = 2$
theories, our results coincide with the results of recent 
\cite{ISnew}. We compare 
the situation in 3d to that in 4d ${\cal N} = 1$ 
theories with massive matter.
In both cases, extra Higgs vacuum states may appear when the Lagrangian 
involves nontrivial Yukawa interactions between the matter superfields. 
In addition, in 3d theories, massive fermion loops affect the index via 
renormalization of the Chern-Simons level $k$.  
\end{abstract}


\end{titlepage}

\section{Introduction}

The Witten index in ${\cal N} = 1,2,3$ supersymmetric Yang-Mills theories
with Chern-Simons term was calculated in \cite{Wit99,N>1,ja1,ja2}. In the 
simplest ${\cal N} =1$ model with the $SU(2)$ gauge group 
(in this paper, we concentrate our attention on $SU(2)$ 
theories, though we will discuss also other 
groups at the end of the paper),
 \be
 \label{LN1}
   {\cal L} \ =\ \frac 1{g^2}   \left \langle - \frac 12 F_{\mu\nu}^2 +
   \lambda/\!\!\!\!\nabla \lambda \right \rangle +
  \kappa    \left\langle \epsilon^{\mu\nu\rho}
  \left( A_\mu \partial_\nu A_\rho - \frac {2i}3 A_\mu A_\nu A_\rho \right ) + i \lambda \lambda \right \rangle
   \ee
 ($\langle \cdots \rangle$ standing for the color trace), the result is
\footnote{If we want the theory to be gauge invariant with respect to large gauge transformation changing the
topological charge of the field, the {\it level} $k$ must be integer.}
 \be
\label{IndpureN1}
I^{\rm SYMCS}_{{\cal N} = 1} \ = k \equiv  4\pi \kappa 
 \ee
  
It can be derived in two ways.

\begin{enumerate}
\item Put the theory in the large spatial box, $g^2L \gg 1$. The problem of counting the vacuum
states in the theory \p{LN1}  can then be reduced to the problem of counting the states in the pure
CS theory with the level renormalized by the fermion loop \cite{Wit99},
 \be
\label{krenorm}
k_{\rm ren}^{{\cal N} = 1} = k_{\rm tree} -  {\rm sgn} ( k_{\rm tree})  \ .
 \ee

\item Put the theory in small box, $g^2L \ll 1$ and count carefully the vacuum states in the effective
Born-Oppenheimer (BO) Hamiltonian \cite{ja1,ja2}. In Sect. 4, we will describe this method in more details.  

\end{enumerate}

A similar result for the ${\cal N} = 2$ theory involving, compared to \p{LN1},
 an extra adjoint matter multiplet 
is \cite{N>1}
  \be
\label{IndN2}
I^{\rm SYMCS}_{{\cal N} = 2} \ =\ |k| - 1 \, .
 \ee

In recent \cite{ISnew}, the index for ${\cal N} = 2$ theories involving extra matter multiplets
was calculated. In the present note, we explain how to do it for 
generic ${\cal N} = 1$ theories
with matter, when working in the language of ${\cal N} = 1$ 3d superfields. For ${\cal N} = 2$ theories, 
we reproduce 
the results of \cite{ISnew}.  
 We give detailed pedagogical explanations concerning their accurate derivation (our method is based on the deformation
${\cal N} = 2\, \to \, {\cal N} = 1$ and is somewhat different
from that in Ref.\cite{ISnew}) and compare the vacuum dynamics of 3d theories 
 with the more familiar 4d situation.

In Section 2, we make a brief review of the vacuum dynamics of 4d SYM theories with matter and 
explain why the index may 
differ from its value in the pure SYM theory even if the matter is nonchiral and massive.  
In Section 3, we describe the 3d SYMCS theories in interest in the ${\cal N} = 1$ superspace approach. 
Section 4 is devoted
to index calculations. 

Before going further, let us clarify the following point. 
In this paper, we are interested in the conventional Witten index.
The latter is well defined only in the theories with mass gap, and that is what we always assume. 
The characteristic mass parameter comes from the constant $1/g^2$ in front of the supersymmetrized Maxwell term. 
On the other hand, a considerable attention has been attracted recently to {\it conformal} 3d supersymmetric CS theories 
because of their remarkable dualities to 11-dimensional supergravities \cite{ABJM}. Witten (alias, toroidal) index 
is not defined in these theories, and the proper tool to study them is the 
so-called superconformal (alias, spherical)
index \cite{Spir}. We will not  touch further upon this issue here.

\section{4d theories.}

We start with reminding what happens in ${\cal N} = 1$ 4d theories. The index of pure SYM theories
was calculated in \cite{Wit82}.   For $SU(N)$ groups
\footnote{We do not discuss here rather nontrivial subtleties in the index calculation for orthogonal and exceptional
groups \cite{Witort,Rosly,jaKac,Keur}},  the result is
 \be
\label{Ind4dSYM}
I_{{\cal N} = 1}^{\rm 4d\ SYM} \ =\ N \, .
 \ee

It was argued \cite{Wit82} that adding {\it nonchiral} matter to the theory does not change
the estimate \p{Ind4dSYM}. Indeed, nonchiral fermions (and their scalar superpartners) 
can be given a mass. For large masses, they seem to decouple and the index seems to be the same as
in the pure SYM theory\footnote{This does not work for chiral multiplets. The latter are always massless and always 
affect the index  \cite{jachiral}.}.  

However, it was realized later that, in {\it some} cases, massive matter 
{\it can} affect the index. The latter may
change when one adds on top of the mass term also Yukawa terms 
coupling different matter multiplets. The simplest example 
\footnote{It was very briefly considered in \cite{ISold} and analyzed in details 
in \cite{GVY}.} is the ${\cal N}= 1$ $SU(2)$ 
theory involving a couple of fundamental matter multiplets $Q^j_f$ ($j = 1,2$ being 
the color and $f=1,2$ the subflavor index; the indices are raised and lowered
with $\epsilon^{jk} = - \epsilon_{jk}$ and  $\epsilon^{fg} = - \epsilon_{fg}$) 
and an adjoint multiplet $\Phi_j^k = \Phi^a (t^a)_j^k$.

Let the tree superpotential be
  \be
\label{massYukawa}
{\cal W}^{\rm tree} = \mu \Phi^j_k 
\Phi^k_j + \frac m2  Q^j_f Q^f_j 
+ \frac h {\sqrt{2}} Q_{jf} \Phi^j_k Q^{kf}  \, ,
  \ee
where $\mu$ and $m$ are adjoint and fundamental masses, and $h$ is the Yukawa constant.

There is also the instanton-generated superpotential \cite{ADS},
 \be
\label{ADS}
{\cal W}^{\rm inst} \ =\ \frac {\Lambda^5}{V} \, ,
 \ee
where $\Lambda$ is a constant of dimension of mass and 
$V =  Q^j_f Q^f_j/2 $ is the gauge-invariant moduli. 
Excluding $\Phi$, we obtain the effective superpotential 
 \be
{\cal W}^{\rm eff} \ =\ mV - \frac {h^2 V^2}{4\mu} + \frac {\Lambda^5}V \, .
 \ee
The vacua are given by the solutions to the equation 
$\partial {\cal W}^{\rm eff} / \partial V = 0$. This equation is cubic, and hence there are 
{\it three} roots and {\it three} vacua.
\footnote{These three {\rm vacua} are intimately related to three {\it singularities} in
the moduli space of the associated ${\cal N} = 2$ supersymmetric theory with a single matter
hypermultiplet studied in \cite{SW94}.}

Note now that, when $h$ is very small, {\it one} of these vacua 
is characterized by a
 very large value,
$\langle V \rangle \approx 2\mu m/h^2$ (and the instanton term in the superpotential plays no role here). 
In the limit $ h \to 0$, it runs to infinity and we are left with only 
{\it two} vacua, the same number as in the pure SYM $SU(2)$ theory. 
Another way to see it is to observe that, for $h = 0$, the equation
  $\partial {\cal W}^{\rm eff} / \partial V = 0$ becomes quadratic having only two solutions.

The same phenomenon shows up in the theory with $G_2$ gauge group studied in \cite{G2}. 
\footnote{ The group 
$G_2$ can be defined as a subgroup of $O(7)$ leaving invariant the structure $f^{jkl} 
A_j B_k C_l$ for any triple of 7-vectors $\vecg{A},\vecg{B},\vecg{C}$, where $f^{jkl} $
is a certain (Fano) antisymmetric tensor.}

This 
theory involves three 7-plets $S^j_f$. The index of a
 pure SYM with $G_2$ group is known to coincide with the 
adjoint Casimir eigenvalue  $c_V$ 
of $G_2$
(another name for it is the dual Coxeter number $h^\vee$). 
It is  equal to 4. 

However, if we 
include in the superpotential the Yukawa term,
 \be
{\cal W}^{\rm Yukawa} \ =\ h\, \epsilon^{fgh} f^{jkl} 
S_{fj} S_{gk} S_{hl} \, , 
  \ee
two new vacua appear. They run to infinity in the limit $h \to 0$. 

The appearance of new vacua when Yukawa  terms are added should by no means come as a surprise. 
This is basically due to the fact that the Yukawa term has higher dimension than the mass term.
Recall that also in the simple non-gauge
Wess-Zumino model, the number of vacua is determined by the power $n$ of the 
superpotential polynomial, $I = n-1$.
\footnote{In 4 dimensions, only the values $n=2,3$ are allowed, 
otherwise the theory is not renormalizable, but one can also think of the dimensionally reduced 
WZ model. If getting rid of all spatial dimensions, $n$ is arbitrary. }

\section{3d gauge theories in 3d ${\cal N} = 1$ superspace.}

The corresponding formalism was developed in \cite{Gates}.
 Our conventions are, however, somewhat different from those in  
\cite{Gates}. For example, 
we prefer vectorial rather than spinorial notations and 
  are using the metric with the signature $(+--)$ rather than $(-++)$.

The superspace $(x^\mu, \theta^\alpha)$ involves a real 2-component spinor $\theta^\alpha$. Indices are
  lowered  and raised with antisymmetric $\epsilon_{\alpha\beta}, \epsilon^{\alpha\beta}$ with the convention
$\epsilon_{12} = - \epsilon^{12} =1$. We define $\theta^2 = \theta^\alpha\theta_\alpha = 2\theta^1\theta^2$ and
$d^2\theta = d\theta^1 d\theta^2$. Then 
 \be
\theta^\alpha \theta_\beta = \frac 12 \theta^2 \delta^\alpha_\beta, \ \ \ 
\theta^\alpha \theta^\beta = -\frac 12 \theta^2 \epsilon^{\alpha\beta}, \ \ \ \ \ 
- \frac 12 \int d^2\theta \, \theta^2 = 1 \, .
 \ee
The 3d $\gamma$-matrices are chosen as 
 \be
\label{gamma}
(\gamma^\mu)^\alpha_{\ \beta} \ =\ (\gamma^0, \gamma^1, \gamma^2)^\alpha_{\ \beta} \ =\ 
(\sigma^2, i\sigma^1, i\sigma^3)^\alpha_{\ \beta} \, .
 \ee
They satisfy the identity
 \be
\label{gamiden}
\gamma^\mu \gamma^\nu \ =\ g^{\mu\nu} + i \epsilon^{\mu\nu\rho} \gamma_\rho
 \ee
with the convention $\epsilon^{120} = 1$. Note that $(\gamma^\mu)_{\alpha\beta}$ are all imaginary and symmetric. The latter implies 
$(\gamma^\mu)^\alpha_{\ \beta} = (\gamma^\mu)_\beta^{\ \alpha}$.

The supersymmetric covariant derivatives are 
\be
{\cal D}_\alpha = \ \frac \partial {\partial \theta^\alpha} + (\gamma^\mu)_{\alpha\beta} \theta^\beta \partial_\mu \, .
 \ee
They satisfy the algebra
 \be
\{{\cal D}_\alpha, {\cal D}_\beta \} = 2(\gamma^\mu)_{\alpha\beta} \partial_\mu \, .
 \ee
Gauge theories are described in terms of the real spinorial superfield $\Gamma_\alpha$. For non-Abelian theories,
$\Gamma_\alpha$ represent Hermitian matrices. As in 4d, one can choose the Wess-Zumino gauge reducing the number
of components of $\Gamma_\alpha$. In this gauge,
 \be
\label{Gamalph}
 \Gamma_\alpha = i(\gamma^\mu)_{\alpha\beta} \theta^\beta A_\mu + i\theta^2 \lambda_\alpha \, ,
 \ee
The covariant superfield strength is
 \be
\label{Walph}
W_\alpha \ =\ \frac 12 {\cal D}^\beta {\cal D}_\alpha \Gamma_\beta - \frac 12 [\Gamma^\beta, {\cal D}_\beta \Gamma_\alpha] \, . 
 \ee
(The full expression \cite{Gates} involves also the term 
$\sim [\Gamma^\beta, \{\Gamma_\beta, \Gamma_\alpha\}]$ but, in the WZ gauge, it vanishes.) $W_\alpha$ 
is expressed into components as
 \be
\label{Wexpan}
 W_\alpha \ =\ -i\lambda_\alpha + \frac 12 \epsilon^{\mu\nu\rho} F_{\mu\nu} (\gamma_\rho)_{\alpha\beta} \theta^\beta
+ \frac {i\theta^2} 2 (\gamma^\mu)^\beta_{\ \alpha} \nabla_\mu \lambda_\beta \, ,
 \ee
where $\nabla_\mu \lambda = \partial_\mu \lambda  - i[A_\mu, \lambda]$, $F_{\mu\nu} = i[\nabla_\mu, \nabla_\nu]$.

In the superfield language, the Lagrangian \p{LN1} is written as
\be
\label{LN1super}
{\cal L} \ =\ \int d^2\theta  \left \langle  \frac 1{2g^2} W_\alpha W^\alpha  + \ \frac {i\kappa}2 
 \left( W_\alpha \Gamma^\alpha + \frac 13 \{\Gamma^\alpha, \Gamma^\beta\}  
{\cal D}_\beta \Gamma_\alpha 
\right) \right \rangle \, . 
 \ee

Let us add now matter multiplets. Consider first the theory with a single real adjoint multiplet,
 \be
\label{Phi}
\Phi = \phi + i\psi_\alpha \theta^\alpha + i\theta^2 D \, .
 \ee
The gauge invariant kinetic term has the form
 \be
\label{LPhikin}
{\cal L}^{\rm kin} = - \frac 1{2g^2}  \int d^2\theta  \left \langle \nabla_\alpha \Phi \nabla^\alpha \Phi  
\right\rangle\, ,
 \ee
where $\nabla_\alpha \Phi = {\cal D}_\alpha \Phi - [\Gamma_\alpha, \Phi]$ and 
the coefficient $-1/(2g^2)$ is chosen for the further convenience. One can add also the mass term,
 \be 
\label{LPhiM}
{\cal L}_M \ =\ -i\zeta  \int d^2\theta \langle \Phi^2 \rangle  \, .
 \ee
Adding together 
\p{LN1super}, \p{LPhikin}, \p{LPhiM},  expressing the Lagrangian in components, and excluding the auxiliary field
$D$, we obtain
  \be
 \label{LN2}
   {\cal L} \ =\ \frac 1{g^2}
  \left\langle - \frac 12 F_{\mu\nu}^2 + \nabla_\mu \phi \nabla^\mu \phi + 
   \lambda/\!\!\!\!\nabla \lambda  +  \psi/\!\!\!\!\nabla \psi \right\rangle \nonumber \\ +
  \kappa   \left\langle \epsilon^{\mu\nu\rho}
  \left( A_\mu \partial_\nu A_\rho - \frac {2i}3 A_\mu A_\nu A_\rho \right ) + i \lambda^2  \right\rangle 
 + i\zeta 
\langle \psi^2 \rangle  - \zeta^2 g^2 \langle \phi^2 \rangle  \, .
   \ee 
The Lagrangian 
 involves, besides the gauge field, the adjoint fermion $\lambda$ with the mass $m_\lambda = \kappa g^2$, the adjoint 
fermion  $\psi$ with the mass $m_\psi = \zeta g^2$ and the adjoint scalar with the same mass. The point $\zeta = \kappa$
is special. In this case, the Lagrangian \p{LN2} enjoys the ${\cal N} = 2$ supersymmetry.

Suppose now that the theory involves {\it two} different adjoint multiplets $\Phi_1$ and $\Phi_2$. 
In this case, we are free to write three different mass terms,
 $$ \sim \int d^2\theta \langle \Phi_1^2 \rangle, \ \ \ \ \ 
\sim \int d^2\theta \langle \Phi_2^2 \rangle, \ \ \ \ \ \ \ \sim \int d^2\theta \langle \Phi_1 \Phi_2 \rangle \, .$$
It is convenient to define the complex combination $\tilde{\Phi} = \Phi_1 + i\Phi_2$ and represent the mass term as
 \be
\label{massrandc}
 {\cal L}_M \ =\ -i \int d^2\theta \left\langle 
\zeta \bar{\tilde{\Phi}} \tilde{\Phi} + \frac 1 2 \left( \rho
 \tilde{\Phi}^2 + \bar\rho  \bar{\tilde{\Phi}}^2 \right)
\right\rangle \, . 
   \ee
One can then call the product $\zeta g^2$ a {\it real} mass $m$ of the complex adjoint multiplet $\tilde{\Phi}$ 
and the product $\rho g^2$ its {\it complex} mass. The complex mass term can also be easily
written in the ${\cal N} =2$ superspace obtained by dimensional reduction from 4d. On the other hand,  the real mass
term can only be written in terms of ${\cal N} = 2$ superfields if introducing extra $\theta$ dependence in the 
integrand \cite{realmass},
 \be
\label{Nishino}
 {\cal L}_{\rm real\ mass} \ \sim \ \int d^4\theta \, 
e^{m \theta \bar\theta} \,\bar{\tilde{\Phi}} \tilde{\Phi} \,.
 \ee
 Such terms modify the standard  ${\cal N} = 2$ superalgebra introducing nonzero central charges. 

In the next section, we will explain why the matter multiplets endowed 
with complex masses (in contrast to those with real masses) do not affect the index 
(up to a possible overall sign flip, which is irrelevant for physics). 

Consider now the theory involving besides the gauge multiplet $\Gamma_\alpha$ a complex fundamental multiplet,
 \be
\label{Qj}
Q_j \ =\ q_j + i\chi_{\alpha j} \theta^\alpha + i F_j \theta^2 \, . 
 \ee
 We add to the gauge Lagrangian \p{LN1super} the terms
  \be
 \label{Lfundsuper}
  {\cal L}^{\rm fund} \ =\ - \frac 1{2g^2} \int d^2\theta  \,  \bar Q^j \nabla^\alpha
\nabla_\alpha Q_j -
i\xi \int d^2\theta \, \bar Q^j  Q_j 
 \ee
[$\bar Q^j = (Q_j)^\dagger,\ \bar \chi^{j\alpha} = (\chi_j^\alpha)^\dagger$, 
$\nabla_\alpha = {\cal D}_\alpha - \Gamma_\alpha$].
After excluding the auxiliary fields $F_j$, this gives in components
 \be
\label{Lfund}
  {\cal L}^{\rm fund} \ =\ -\frac 1{g^2} ( \bar q^j  \nabla^\mu \nabla_\mu q_j + m^2 \bar q^j q_j) 
+   \frac 1{g^2} ( \chi_j /\!\!\!\!\nabla  \bar \chi^j  + im \bar \chi^j \chi_j ) 
 \ee
with $m = \xi g^2$. 

If two different fundamental multiplets are added, one can write on top of the real mass term in \p{Lfundsuper} 
also the complex mass term.

Note now that 
 the {\it free} kinetic term in \p{Lfundsuper} (with $\nabla_\alpha \to {\cal D}_\alpha$)
  enjoys in fact the ${\cal N} = 2$ supersymmetry 
(when also the real mass term $\propto \xi$ is included, it is deformed by central charges). 
Indeed, it can be written in terms
of a {\it chiral} ${\cal N} = 2$ superfield $\tilde{Q}_j$ as $\int d^4\theta \, \bar {\tilde Q}^j  {\tilde Q}_j 
 $.

 The paper \cite{ISnew} was devoted to calculating the index in interacting  ${\cal N} = 2$ theories with 
central charges. The simplest such (non-Abelian) theory involves the  ${\cal N} = 2$ gauge multiplet 
and a fundamental matter multiplet. The  ${\cal N} = 2$ symmetric Lagrangian represents the sum of 
the terms like in \p{LN1super}, \p{LPhikin}, and \p{LPhiM} for the gauge multiplet, the kinetic and the mass terms 
\p{Lfundsuper} for the fundamental matter and, on top of that, the Yukawa term \cite{Ivanov}
 \be
\label{3dYukawa}
  {\cal L}_{\rm Yukawa} \ =\ \frac i {g^2} \int d^2\theta \, \bar Q^j \Sigma_j^{\ k} Q_k  \, .
 \ee  
(We have renamed here the adjoint multiplet, $\Phi \to \Sigma$, to facilitate the comparison with Ref.\cite{ISnew}.)

Likewise, one can consider the  ${\cal N} = 2$ SYMCS theory coupled to the complex adjoint  ${\cal N} = 2$
multiplet $\Phi$ endowed with a real mass. The Lagrangian includes an extra Yukawa term
 $\propto \int d^2\theta \, \langle \Sigma  \Phi \bar \Phi \rangle$. 

We will see that, in the theories involving Yukawa terms and, in particular, in  ${\cal N} = 2$ theories, extra 
vacuum states on the Higgs branches appear by the same mechanism as 
   in 4d theories.    

\section{Index calculations.}

\subsection{ Pure ${\cal N} = 1$ SYMCS theory.}


Let us first remind how the result \p{IndpureN1} is derived in the BO approach.

\begin{itemize}

\item Put the theory in a small spatial box, $g^2L \ll 1$, and impose 
{\it periodic} boundary conditions on all fields.
\footnote{We stick to this choice here though, in a theory 
involving only adjoint fields, one could also impose the so-called 
{\it twisted} boundary conditions. In 4d theories, this results 
in the same value for the index \cite{Wit82}, 
but, in 3d theories, the result turns out to be 
different \cite{Henningson}.} 

\item The effective BO Hamiltonian involves slow variables, which in this case are just the zero Fourier modes
of the spatial components of the Abelian vector potential and its superpartners,
 \be
\label{Cj}
C_j \ =\ A_j^{({\bf 0}) 3}, \ \ \ \ \ \ \ \ \ \  \lambda_\alpha \ =\ 
\lambda_\alpha^{({\bf 0}) 3} \, . 
 \ee
(Here $j=1,2$ is  the spatial index.)
\item Note that the shift 
 \be
 \label{shift}
C_j \to C_j + 4\pi/L
 \ee
 amounts to a  gauge transformation. Gauge invariance then dictates for the BO wave functions to satisfy 
certain boundary conditions. In 4d theories, the effective wave functions should simply be periodic under the shift 
\p{shift}. In 3d theories, they are periodic modulo certain phase factors \cite{DJT,ja1},
 \be
\label{bc}
\Psi(X+1,Y) &=& e^{-2\pi i kY} \Psi(X,Y) \, , \nonumber \\
\Psi(X,Y+1) &=& e^{2\pi i kX} \Psi(X,Y) \, ,
  \ee
 where $X = C_1 L/(4\pi),  Y = C_2 L/ (4\pi) $. 

\item At the tree level, the effective Hamiltonian describes the $2d$ motion in a homogeneous magnetic field,
\be
\label{Heff}
H^{\rm eff} \ =\ \frac {g^2}{2L^2} \left( P_j - \frac {\kappa L^2}{2} \epsilon_{jk} C_k \right)^2 + \frac {\kappa g^2}2
(\lambda \bar\lambda - \bar \lambda \lambda) \, ,
 \ee
where $\lambda = \lambda_1 -i\lambda_2, \bar \lambda = \lambda_1 + i\lambda_2$. 
For positive $\kappa$, the ground states of this Hamiltonian are bosonic. For negative $\kappa$, they are fermionic. 

Were the motion on the plane $(C_1, C_2)$ infinite, the ground state would be infinitely degenerate. But 
the presence of the boundary conditions \p{bc} implies that the motion is finite, with $C_j$ lying on the {\it dual torus}
 of size $4\pi/L$. The level of degeneracy is then determined \cite{Novikov} by the magnetic flux on the dual torus, which is equal to 
$2k$ in this case. The eigenfunctions of the vacuum (and all other) states can be written explicitly. 
They are expressed via elliptic $\theta$ functions. 

\item Note now that not all $2|k|$ states are admissible. We have to impose the additional Weyl invariance condition (following
from the gauge invariance of the original theory). For $SU(2)$, this amounts to 
\footnote{Note that, in contrast to what should be done in 4 dimensions \cite{Wit82},  we did not include here the Weyl reflection
of the fermion factor $\lambda$ entering the effective wave function for negative $k$. The reason is that the conveniently 
defined {\it fast} wave function (to which the effective wave function depending only on $C_j$ and $\lambda$ should be multiplied) 
involves, for negative $k$, a Weyl-odd factor $C_1+iC_2$. This oddness compensates the oddness of the factor $\lambda$ in the effective
wave function \cite{ja1}.}  
$\Psi^{\rm eff}(-C_j) = 
 \Psi^{\rm eff}(C_j )$, which singles out $|k| + 1$ vacuum states, bosonic for $k > 0$ and fermionic for $k < 0$.

When $k=0$, the effective Hamiltonian \p{Heff} describes free motion on the dual torus. There are two zero energy 
ground states, 
$\Psi^{\rm eff} = {\rm const}$ and   
$\Psi^{\rm eff} = {\rm const}\cdot \lambda$ 
(we need not to bother about Weyl oddness of the factor $\lambda$ 
by the same reason as above). The index is zero. We thus derive
 \be
\label{Itree}
I^{\rm tree} \ =\ (|k| + 1){\rm sgn} (k) \, .
 \ee

\item
However, the expression \p{Itree} is not the correct result for the index yet. 
One has to take into account loop corrections. They are negligible in the bulk of the dual torus, but modify 
the effective Hamiltonian essentially at the vicinity of four special points (the ``corners''), 
\be 
\label{corners} 
C_j = 0, \ C_j = (2\pi/L, 0), \ 
C_j = (0, 2\pi/L), \ C_j = (2\pi/L, 2\pi/L) \, , 
 \ee
where the ``Abelian'' BO approximation breaks down.

Consider for definiteness the case of positive $k$. 
There are corrections coming from the gluon loops and from the fermion loops. 
We explore the theory in the limit $g^2L \ll 1$ when the fermion and gluon mass $m = \kappa g^2$ is 
much smaller than
the size of the dual torus $\sim 1/L$. When the mass is disregarded altogether, {\it gluon} loops bring about
the $\delta$-singular flux lines with unit flux +1 (a kind of Dirac strings) in each corner. 
These lines are unobservable. An accurate analysis \cite{ja2,ja3} shows that the corrections coming from the 
gluon loops can be disregarded also when a small finite mass is taken into account. 
As for the fermion loops, they bring about the vortices with the {\it fractional} fluxes  
$\Phi_{\rm corner} = - \frac 12$. The net fermion-induced flux is integer, $\Phi_{\rm induced} = -2$.

The rule of thumb (see \cite{ja2,ja3} for more details) is that the vacuum states are counted
correctly if tree-level $k$ is renormalized by 
the fermion loops only, 
leading to \p{krenorm} and to \p{IndpureN1}. The states associated with gluon flux lines might be present for 
finite $m$ in the effective Abelian BO Hamiltonian, but they do not correspond
to admissible states with nonsingular wave functions 
in the full Hamiltonian.

 Two reservations are of order.   {\it (i)}
\p{IndpureN1} follows smoothly from \p{Itree} and \p{krenorm} only when $|k| > 1$.
When e.g. $k^{\rm tree}=1$, Eq.\p{krenorm} gives $k_{\rm ren}= 0$ and, if substituting this in \p{Itree}, 
one should assume sgn$(0)$ =1 rather than   sgn$(0)$ =0, which one should assume for 
$k^{\rm tree} = k_{\rm ren} = 0$. {\it (ii)}
 Also for $|k| > 1$, the formula \p{krenorm} should be understood
{\it cum grano salis} because, in contrast to the homogeneous 
tree-level effective magnetic field on the dual torus, the loop-induced one  is singular at the corners.

At any rate, the number of vacuum states can be determined using the  shifted flux (\ref{krenorm}). 
If $k >0$, the effective wave functions 
represent Weyl-invariant combinations ($k$ of them) of the functions 
  \be
\label{koren}
 \Psi^{\rm eff}(X,Y) \  \sim \  
 Q^{2k-2}_m(\bar z) \Pi^{3/4} (\bar z) \Pi^{-1/4} (z) \, ,
 \ee
where $z = X + iY$, $Q^{2k-2}_m(\bar z)$ 
are $\theta$ functions of level $2k-2$, and   
 $\Pi(z)$ is a certain $\theta$ function  of level 
4 having zeros  at  $ z = 0,
1/2, i/2, (1+i)/2$  corresponding to the corners \p{corners}.
\footnote{The function $\Pi(z)$ is known from the studies of canonical quantization of pure CS theories 
\cite{Pi}. }  The functions \p{koren}  vanish 
at the corners.    

The effective wave functions at negative $k$ have a similar form, one has only to 
interchange $z$ and $\bar z$ and add
the extra fermionic factor $\lambda$.

\end{itemize}

The result \p{IndpureN1} implies spontaneous supersymmetry breaking 
in the pure SYM theory with $k=0$.
\footnote{To be on the safe side, one should have rather said {\it suggests} instead
of {\it implies}, the vanishing of the index is a necessary but not sufficient condition
 for the spontaneous supersymmetry breaking. However, in most cases when there are no special reasons to the 
contrary
(like the presence of an extra symmetry and an extra nonvanishing associated index \cite{Wit82}), 
supersymmetry breaks
if $I_W = 0$. 
We will assume that this happens in all theories with vanishing index discussed in this paper.}

We want to emphasize that the result \p{IndpureN1} for the index refers to 
the theory \p{LN1} involving besides the Chern-Simons term also the 
Maxwell term. In a {\it pure} supersymmetric CS theory, fermions are 
not coupled to the gauge fields, there is no renormalization \p{krenorm}, 
and the index is given by \p{Itree} rather than by \p{IndpureN1}. 
 


\subsection{ ${\cal N} = 1$:   adjoint matter.}


Let us consider now the theory \p{LN2} with a single extra adjoint matter multiplet. Let first $\zeta > 0$. Then the mass of the 
matter fermions is positive. To be more precise, it has the same sign as the gluino mass for $k > 0$. The matter loops  bring
about an extra  renormalization of $k$.

Note that the status of this renormalization is different compared to that due to the gluino loop. As was mentioned, for the latter, 
the induced magnetic
field on the dual torus is concentrated in the corners \p{corners}, which follows from the equality $m_\lambda L \ll1$. 
 On the other hand, the mass of the matter fields $m_\psi = \zeta g^2$ is an independent parameter. 
It is convenient to make it {\it large}, $m_\psi L \gg 1$. For a finite mass, the induced magnetic field has the form \cite{ja1}
  \be
\label{BindC}
\Delta {\cal B} (\vecg{C}) \ =\ - \frac {m_\psi} 2 \sum_{\vecg{n}} \frac 1 {\left[ \left( \frac {2\pi \vecg{n}}L - \vecg{C} \right)^2 +
m_\psi^2 \right]^{3/2}}
   \ee
For small $m_\psi L$, it is concentrated in the corners. But in the opposite limit, the induced flux density becomes constant, as the
tree flux density is. 

\begin{figure}[ht!]
     \begin{center}

            \includegraphics[width=0.6\textwidth]{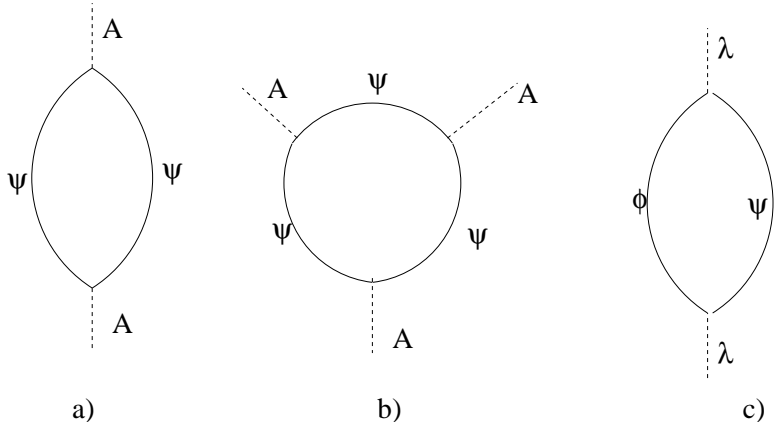}
        
    \end{center}
    \caption{ Renormalization of $k$ due to matter loops: (a), (b) Chern-Simons term;  (c)  gluino mass term.}
\label{graphs}
    \end{figure}

Thus, massive matter brings about a true renormalization of $k$ without any 
 qualifications ({\it sine sale} if you will).
Note that the gluino mass term
in \p{LN1} also gets renormalized. The graphs responsible for the renormalization of the bosonic CS structure and the term $\sim 
\lambda \lambda$ are depicted in Fig.\ref{graphs}.    

For positive $\zeta$, the renormalization is negative, $k  \to  \ k-1$. The index coincides with 
the index of the ${\cal N} = 1$ SYMCS theory with the renormalized $k$,
  \be
\label{poszeta}
 I_{\zeta > 0}  = k -1 \, .
 \ee
For $k = 1$, the index is zero and supersymmetry is spontaneously broken.

For negative $\zeta$, two things happen. 

\begin{itemize}

\item
First, the fermion matter mass has the opposite sign and so does the renormalization of $k$ due
to the matter loop. We seem to obtain $I_{\zeta < 0} =  k + 1$.

\item This is wrong, however, due to another effect. For positive $\zeta$, the  ground state wave function 
in the matter sector is bosonic. But for negative $\zeta$, 
it is fermionic, $\Psi \propto \prod_a \psi^a$, changing the sign of the index. 

\end{itemize}

 We obtain
 \be
\label{negzeta}
 I_{\zeta < 0} = -k - 1 \, .
 \ee
Supersymmetry is broken here for $k = -1$. 

As was mentioned, the Lagrangian \p{LN2} with $\zeta = \kappa$ enjoys 
the extended ${\cal N} = 2$ supersymmetry. That means, in particular,
that $\zeta$ changes the sign together with $\kappa$ and the result 
is given by \p{IndN2}. The latter expression [in contrast to
\p{poszeta} and \p{negzeta}] is not analytic at $k=0$, this nonanalyticity being  due just to  the sign flip 
 of the matter fermion mass. 
Strictly speaking, 
the formula \p{IndN2} does not work for $k=0$. In this case, 
also $\zeta = 0$, the matter is massless, massless scalars make the
motion infinite and the index is ill-defined. However, bearing 
in mind that the regularized theory with $\zeta \neq 0$ gives the 
result  $I^{\rm SYMCS}_{{\cal N}= 2\ {\rm deformed}}(0) = -1$, 
irrespectively of the sign of $\zeta$, one can attribute this value for the index also to
$ I^{\rm SYMCS}_{{\cal N}= 2 }(0)$.

\begin{figure}[ht!]
     \begin{center}

            \includegraphics[width=0.6\textwidth]{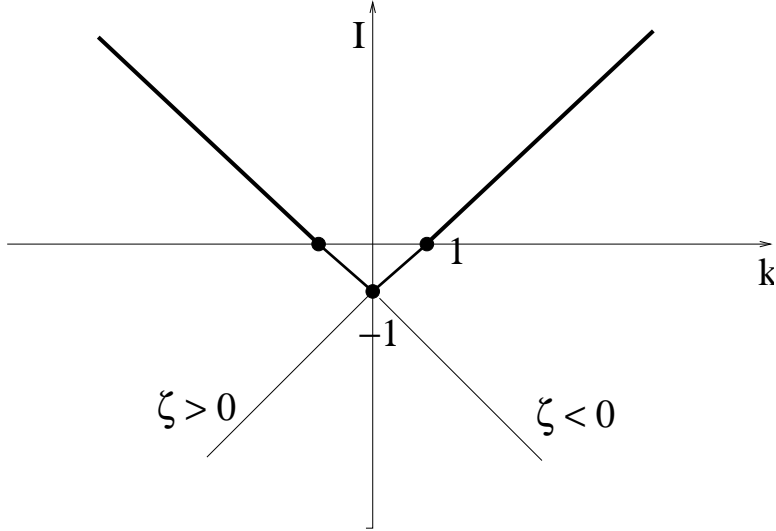}
        
    \end{center}
    \caption{ The indices in the theory 
\p{LN2} with $\zeta > 0$, $\zeta < 0$, and $\zeta = \kappa$ (bold lines). }
\label{triind}
    \end{figure}

The three index formulas \p{poszeta}, \p{negzeta}, and \p{IndN2} are represented together in Fig.\ref{triind}.

Let now the theory involve two extra real or one extra complex
 adjoint matter multiplet. As was discussed in Sect. 3, the matter fields 
can in this case be endowed with a real mass or with a complex mass 
[see Eq.\p{massrandc}]. It is important to understand that 
this choice affects the value of the  index. 

If the mass is real, 
it just means twice as large renormalization of $k$. For positive
real mass, $\zeta > 0$, the index reads
 \be
\label{Ind2adj+}
I^{\zeta > 0}_{\rm two\ adjoint\ multiplets} \ =\ k-2 \, .
 \ee
and, for the negative mass, it is
 \be
\label{Ind2adj-}
I^{\zeta < 0}_{\rm two\ adjoint\ multiplets} \ =\ k+2 \, .
 \ee

Note that, in this case, the extra factor $-1$ is absent. 
We have two matter multiplets now and, for $\zeta < 0$, the fast ground state
wave function includes two fermionic factors and stays bosonic.

If $\zeta =0$ and we have only complex mass $\mu = \rho g^2$ at our disposal, 
$k$ is not renormalized whatsoever. One of the ways to see that is to choose
 $\rho$ to be real.     The mass term \p{massrandc} is reduced 
in this case to $\int d^2\theta [\langle \Phi_1^2 \rangle - 
\langle \Phi_2^2 \rangle ]$, i.e. the masses of $\Phi_1$ and 
$\Phi_2$ have opposite signs, and the associated renormalizations are also
opposite. On the other hand, the fast ground state wave function is now {\it fermionic}. 
Thus, the result for the index coincides with \p{IndpureN1} up to an irrelevant for physics sign flip,
  \be
I_\rho \ = \ -k  \, .
  \ee


\subsection{ ${\cal N} = 1$:  fundamental matter.}


Consider now the ${\cal N} = 1$ theory involving an extra fundamental multiplet with the Lagrangian \p{Lfund}. 
Again, the matter fermion loops affect $k$. The shift of $k$ is 
half as much as in the adjoint case.
\footnote{When calculating the adjoint real fermion loop, the color and 
reality give the factor $c_V \delta^{ab}/2 = \delta^{ab}$. 
For the complex fundamental loop, the factor is 
Tr$\{t^a t^b\} = (1/2) \delta^{ab}$.} There are two fermion components $\chi_1, \chi_2$ and the ground state wave function in the matter sector 
is  bosonic, irrespectively of the sign of the mass. We obtain,
  \be
  I = \ k - \frac 12 {\rm sgn}(\xi) \, .
   \ee

Note that, for consistency, $k$ should be half-integer here. This can be explained, if observing that the
 large gauge transformations 
(see the footnote 2) not only add here $2\pi i k$ to the Minkowski action, but also change the 
sign of the fermion determinant in the
functional integral \cite{Kao2Lee}. 
The same refers to pure ${\cal N} = 1$ SYMCS theories with higher groups. For example, for $SU(N)$, $k$ should
be integer when $N$ is even and half-integer when $N$ is odd.


\subsection{ ${\cal N} = 2$:  fundamental matter.}


Following the logics of \cite{ISnew}, let us discuss now the ${\cal N} = 2$ theory involving the gauge and 
fundamental matter multiplets. The latter is  endowed with a real mass. As was mentioned before, being expressed in terms 
of    ${\cal N} = 1$ superfields, its Lagrangian reads
 \be
\label{fundN2}
{\cal L} = \ \int d^2\theta  \left \langle  \frac 1{2g^2} W_\alpha W^\alpha  + \ \frac {i\kappa}2 
 \left( W_\alpha \Gamma^\alpha + \frac 13 \{\Gamma^\alpha, \Gamma^\beta\}  
{\cal D}_\beta \Gamma_\alpha 
\right) \right \rangle  \nonumber \\
- \frac 1{2g^2}  \int d^2\theta  \left \langle \nabla_\alpha  \Sigma \, \nabla^\alpha \Sigma 
\right\rangle   -i\kappa  \int d^2\theta \langle \Sigma^2 \rangle 
- \frac 1{2g^2} \int d^2\theta  \,  \bar Q^j \nabla^\alpha
\nabla_\alpha Q_j -
i\xi \int d^2\theta \, \bar Q^j  Q_j \nonumber \\
+ \frac i {g^2} \int d^2\theta \, \bar Q^j \Sigma_j^{\ k} Q_k \, .
  \ee  

To calculate the index, we {\it deform} the theory substituting for $\kappa$ 
in the second line some large constant $\zeta$. ${\cal N} = 2$ supersymmetry is 
then broken down to ${\cal N} = 1$, but the index is the same as before. One should only take care that the sign of $\zeta$
is the same as the sign of $\kappa$, to avoid passing the singularity at $\zeta = 0$.
 
In a deformed theory, the mass of the multiplet $\Sigma$ is  $M = \zeta g^2$.
The mass of the fundamental multiplet is $m = \xi g^2$. We assume both of 
them to be large, $ML \sim mL \gg 1$. (As we always keep $g^2L$ small, 
this means also $M \sim m \gg g^2$ and $\zeta \sim \xi \gg 1$.) Then 
  $k$ is renormalized by fermion loops with quasi-homogeneous flux densities. We are thus in a position 
to evaluate the index of the pure ${\cal N} = 1$ theory with renormalized $k$.

There are four different cases:
\footnote{When comparing with \cite{ISnew}, note the mass sign convention for the matter fermions is
 {\it opposite} there compared to our convention. We call the mass positive if it has the same sign as the 
masses of fermions in the gauge multiplet for positive $k$ (and hence positive $\zeta$). In other words, for 
positive $k,\xi$, the shifts of $k$
due to both adjoint and fundamental fermion loops have the negative sign.}
 
\begin{enumerate}

\item $m > 0, \ \ k>0 \Rightarrow M > 0$.
\be
\label{krenSU2}
k  \to  k -  1_{\rm adj.\ matter} - 
\left(\frac 12\right) _{\rm fund.\ matter} = k- \frac 32 \, .
 \ee
This contributes $k - \frac 32$ to the index. Note that, when $k= \frac 12$, this contribution is negative.

\item $m > 0, \ \ k,M < 0$.
 \be
k  \to  k  + 1_{\rm adj.\ matter} - 
\left(\frac 12\right)_{\rm fund.\ matter} = k + \frac 12 \, .
 \ee
Multiplying it by -1 due to the fermionic nature of the wave 
function in the adjoint matter sector
[see the discussion before Eq.\p{negzeta}], we obtain $I = -k - 1/2$.

\item $m < 0, \ \ k,M >0$.
 \be
 k \to  k  - 1_{\rm adj.\ matter} + 
\left(\frac 12\right)_{\rm fund.\ matter} = k- \frac 12 \, ,
 \ee
giving the contribution $I = k-1/2$.

\item $m < 0, \ \ k,M<0$.

 \be
k  \to  k  + 1_{\rm adj.\ matter} + \left(\frac 12\right)_{\rm fund.\ matter}
 = k+ \frac 32 \, .
 \ee
The contribution to the index is $ -k - 3/2$.

\end{enumerate}

Note that there is no overall change of sign for negative $\xi$. 
because of the presence of two ${\cal N} =1$ matter multiplets [ see the comment after Eq.\p{Ind2adj-}].
 Note also that we did not include here the renormalization \p{krenorm} due to the gluino loop. 
It is already taken
into account in \p{IndpureN1}.

 For the time being, we have
 \be
\label{IndN2fundbezYuk}
m > 0:  \ \ \ \ I \ =\ \left\{ 
\begin{array}{c}
k-\frac 32, \ \ \  k > 0 \\   -k - \frac 12, \ \ \ k<0 
\end{array} \right. 
 \nonumber \\
m < 0:  \ \ \ \ I \ =\  \left\{ 
\begin{array}{c}
k - \frac 12, \ \ \ k > 0 \\  - k - \frac 32, \ \ \  k <0   
\end{array} \right.  \, .
  \ee 

This is not yet, however, the end of the story.
As we mentioned before, the presence of the Yukawa term in \p{fundN2} 
may lead to appearance of extra vacuum states on the Higgs branch. In the half of the cases listed above, it does.

The component bosonic potential following from \p{fundN2} reads
\be
\label{bospotaux}
V \ =\  -\frac 2{g^2} (D^a)^2 + 2 \zeta  \sigma^a D^a 
- \frac 4{g^2} \bar F F +
2\xi (\bar F q + \bar q F) - \frac 2{g^2} \left( \sigma^a \, \bar F
 t^a q + 
 \sigma^a \, \bar q t^a  F + D^a \, \bar q t^a q \right) 
 \ee
[$\sigma_j^{\ k} = \sigma^a (t^a)_j^{\ k},\ \ 
D_j^{\ k} = D^a (t^a)_j^{\ k}$].
 Excluding the auxiliary fields, we obtain
  \be
 \label{bospot}
g^2 V =  ( m \bar q - \sigma^a \, \bar q t^a)(mq - \sigma^a \, t^a q) 
+ \frac 12 \left(M \sigma^a -  \bar q t^a q \right)^2 \, .
 \ee
When $M \sim m \gg g^2$, this is not renormalized by loops.
The potential vanishes when
 \be
\label{uslovmin}
 m q &=& \sigma^a t^a q \nonumber \\
M \sigma^a &=& \bar q t^a q \, .
 \ee

The equations \p{uslovmin}  have a trivial solution $\sigma = q = 0$, but there is also a nontrivial one. By a gauge rotation, one
can always assure $\sigma^a = \sigma \delta^{3a}$ with positive $\sigma$. Let $m > 0$. Then the first equation in \p{uslovmin}
implies $q_2 = 0$ and the second gives $2M\sigma = |q_1|^2$. This has a solution when $M > 0$, i.e. $k > 0$. (The phase of $q$ can be
 unwinded, of course, by a gauge transformation.) Similarly,
when $m<0$, it is $q_1$ that vanishes and the solution exists for negative $k$ and $M$.

Note that the $SU(2)$ gauge symmetry is broken completely at this minimum. No light fields are left, 
there is no BO dynamics and a classical vacuum corresponds to a single quantum state. 
 
Adding when proper this extra (bosonic) state to the index 
\p{IndN2fundbezYuk}, we obtain the
final universal result \cite{ISnew}.
  \be
 \label{IndN2fund}
 I^{{\cal N} = 2}_{\rm one\ fund.\ mult.}  \ =\ |k| - \frac 12 \, .
  \ee

Supersymmetry is broken for $|k| = 1/2$. One can observe that the modification compared to \p{IndN2} is  minimal here.
Basically, the change $|k| - 1 \to |k| - 1/2$ reflects the fact that $k$ has to be half-integer now rather than integer.

One can  compare the situation with what happens in 4d and note that

{\it (i)}  In four dimensions, to generate an extra Higgs vacuum, one needs a complex adjoint matter multiplet and at least two fundamentals.
In 3d, one can write the Yukawa term, like in \p{fundN2}, for a single ${\cal N}=2$ fundamental multiplet and a real adjoint ${\cal N}=1$ 
multiplet.
This turns out to be  sufficient for the extra state to appear.

 {\it (ii)} This all (both the renormalization of $k$ due to fermion loops and the appearance of the extra Higgs vacuum) 
depends crucially on the presence of the real mass term.
 With a single matter multiplet, we have no choice:
one cannot ascribe it a complex mass and, when all masses are zero, the index is ill-defined.  
 But in a theory with {\rm two} matter multiplets $Q_j^f$ ,  $f = 1,2$, 
one can set real masses to zero and introduce only the complex mass term
\be
\label{complmass}
 {\cal L}_M \ =\ -i \frac \rho 2 \int d^2\theta \, Q^j_f Q_j^f \ + {\rm c.c.} \, . 
   \ee
(such that ${\cal N} = 2$ supersymmetry is not deformed). 

As we noticed at the end of Sect. 4.2, 
complex mass does not renormalize $k$. In addition, no extra Higgs vacua are 
generated. 
Indeed, as one can easily derive, the bosonic potential would vanish in this case provided
   \be
\label{uslovmin2}
 M\sigma = \bar q^f t^3 q_f \, , \nonumber \\
 \mu  \bar q_{fj} + \sigma (t^3 q)_{fj} = 0 
   \ee
with $\mu = \rho g^2$. In contrast to \p{uslovmin}, these equations do not have nontrivial solutions. 
The answer for the index is hence the same as in the
pure ${\cal N} = 2$ SYMCS theory, $I = |k|-1$. (No sign flip here.)

On the other hand, when the theory is regularized with real masses (the same for both real multiplets), the index is \cite{ISnew}
\be
 \label{IndN2fund2}
 I^{{\cal N} = 2}_{\rm 2\ fund.\ mult.,\ real\ masses}  \ =\ |k| \, .
  \ee

The supersymmetry is thus broken for $|k| = 1$ in the theory with complex masses and stays intact in the theory with real masses.


\subsection{ ${\cal N} = 2$: adjoint matter.}


Consider now the ${\cal N} = 2$ theory with  a  complex
 adjoint matter multiplet. 
We can give it a real or a complex mass.
 Let first the mass $m$ be real. It brings about the renormalization of $k$. $k$ is also renormalized due to
the real ${\cal N} = 1$ adjoint multiplet from the  ${\cal N} = 2$ gauge multiplet.  
One can repeat the same analysis as we did in the fundamental case (in particular, we deform the model 
by attributing a large mass $M$ to the real
multiplet $\Sigma$) to obtain the 
following contributions to the index,

\be
\label{IndN2matt}
I^{{\cal N} = 2}_{\rm gauge \ +\ adjoint\ matter} \ =\ \left\{ 
\begin{array}{c}
k-3, \ \ \  k > 0 \\   -k+1, \ \ \ k<0 
\end{array} \right. 
  \ \ \ \ \ \ {\rm if} \ m  > 0 \nonumber \\
I^{{\cal N} = 2}_{\rm gauge\ +\ adjoint\ matter} \ =\  \left\{ 
\begin{array}{c}
k+1, \ \ \ k > 0 \\  - k - 3, \ \ \  k < 0  
\end{array} \right.
  \ \ \ \ \ \ {\rm if} \ m <  0 \, .
  \ee

As in the fundamental case, this is not the full answer yet. 
There are also additional states on the Higgs branch that contribute.
The conditions for the bosonic potential to vanish are the same as in \p{uslovmin}, with the adjoint generators
being substituted for the fundamental ones. We obtain
 \be
\label{uslovminadj}
m\phi^b\  = \ i\epsilon^{abc} \sigma^a \phi^c \, , \nonumber \\
 M\sigma^a \ =\ i  \epsilon^{abc}  \bar \phi^b \phi^c \, .
 \ee
 These equations have  nontrivial solutions when both $M$ and $m$ are positive or when both $M$ and $m$ are negative. Let them be positive.
Then one of the solutions to \p{uslovminadj} is
   \be
\label{soladj}
   \sigma^a \ = \  m\delta^{a3}, \ \ \ \ \ \ \phi = \sqrt{\frac {Mm}2} 
\left( \begin{array}{c} 1 \\ -i \\ 0 \end{array} \right)\, . 
 \ee

At this point, a new important effect comes into play. 
In contrast to the fundamental case where a similar classical solution gave a {\it unique} vacuum state, 
we obtain here {\it four} new states. Indeed, besides the solution \p{soladj}, 
there are also the solutions obtained
from that by  gauge transformations. The latter are not necessarily global, 
they might depend on the spatial coordinates $x,y$.
[Do not confuse them with the dual torus coordinates $X,Y$ introduced after Eq.\p{bc}.]
Note now that, for the theory defined on a torus, one can also apply to \p{soladj} some  
transformations which look like gauge
transformations, but are not  contractible due to the nontrivial $\pi_1[SO(3)] = Z_2$\footnote{One 
should understand $SO(3)$ here not as the orthogonal group itself, but rather as the  adjoint representation space. See the discussion of higher isospins below.}
An example of such a quasi-gauge transformation is 
  \be
\label{largegauge}
\Omega_1: O^{ab}(x) \ =\ \left( \begin{array}{ccc} 
\cos \left( \frac {2\pi x}L \right) &  \sin \left( \frac {2\pi x}L \right) & 0 \\
-\sin \left( \frac {2\pi x}L \right) & \cos \left( \frac {2\pi x}L \right) & 0 \\
 0 & 0& 1   \end{array}   \right) \, ,
  \ee
where $L$ is the length of our box. The transformation \p{largegauge} does not affect 
$\sigma^a = \sigma \delta^{a3}$ and keeps the fields 
$\phi^a(\vecg{x})$ periodic. Note that, for the matter in fundamental representation, the transformation 
\p{largegauge} is inadmissible:
when lifted up to $SU(2)$, it would make a constant solution of \p{uslovmin} antiperiodic. 
There is a similar transformation $\Omega_2$ along the second cycle of the torus.

In 4d theories, wave functions are invariant under contractible gauge  transformations. In 3d SYMCS theories, they
are invariant up to a possible phase factor, like in \p{bc}. 
But nothing dictates the behaviour of the wave functions under the transformations $\Omega_{1,2}$. The latter are actually {\it not}
gauge symmetries, but rather some global symmetries of the theory living on a torus. We obtain thus four different wave 
functions, even or odd under the action of $\Omega_{1,2}$. 
\footnote{The oddness of a wave function under the transformation \p{largegauge} means  nonzero {\it electric flux} in the language
of Ref. \cite{Hooft}.}  
 
 The final result for the index of the theory regularized with the real masses is
\be
 \label{IndN2adj2}
 I^{{\cal N} = 2}_{\rm compl.\ adj.\ mult.,\ real\ masses}  \ =\ |k| + 1 \, .
  \ee

On the other hand, when the theory is regularized with complex masses, the presence of the matter 
has no significant effect on the index,
 and the result \p{IndN2} is left intact up to a sign flip due to the fermion nature of the fast ground state wave function.
 This refers in particular to ${\cal N} = 3$ SYMCS theories
\cite{N>1,ja1}.

The result \p{IndN2adj2} as well as \p{IndN2} was derived under the condition $k \neq 0$. Otherwise, the adjoint scalars in the
 multiplet $\Sigma$ 
become massless. But, similarly to the case of pure ${\cal N} = 2$ discussed above, one can regularize the theory by adding a small  
term $\mu \int d^2\theta \, \langle \Sigma^2 \rangle $  to the Lagrangian
The index is given then by Eq.\p{IndN2adj2}, irrespectively of the sign of $\mu$.

The result   \p{IndN2adj2} (as well as \p{IndN2fund}) 
was  derived in \cite{ISnew} following a different logic. Intriligator and Seiberg 
did not deform  ${\cal N} = 2 \ \to {\cal N} = 1 $ and kept the fields in the  
real adjoint matter multiplet $\Sigma$ light. Then the light matter fields $\{ \sigma, \psi\}$ 
 enter the effective BO Hamiltonian at the same ground as the Abelian components of the gluon and gluino fields.
 As we mentioned,
the fluxes induced
by the light fields are not homogeneous being concentrated at the corners. 
This makes an accurate analysis  essentially more difficult.
The index \p{IndN2adj2} was obtained in \cite{ISnew} as a sum of 
{\it three} rather than just two contributions 
\footnote{On top of the usual vacua with $q = \sigma =0$ and the Higgs vacua with $q,\sigma \neq 0$, 
they had also ``topological vacua'' with $q=0, \sigma \neq 0$. The latter do not appear in our approach. }
and it is still not quite clear  how it works  in the particular case $k=2$ where
$k_{\rm eff}$ as defined in  Ref.\cite{ISnew} and including only renormalizations 
due to complex matter multiplet, $k_{\rm eff} = k-2$,
vanishes. 



\subsection{ ${\cal N} = 2$:  generic matter content.}


When the ${\cal N} = 2$ SYMCS theory is coupled to a complex matter multiplet with an arbitrary isospin $I$ 
endowed with a real mass, 
the index \p{IndN2} is shifted up by \cite{ISnew}
  \be
\label{Dynkin} 
 \frac 12 T_2(I) \ = \frac {I(I+1)(2I+1)} 3 
  \ee
[with $T_2(I)$ standing for the Dynkin index of the corresponding representation normalized to 
$T_2$({\sl fund}) $=1$]. 
 When deriving this, one should take into account the renormalization of $k$ and add the Higgs vacua. 

  Let us make a brief comment on how the
latter are counted taking $I = 3/2$ as an example. We have again the equations \p{uslovmin} 
with the generators $T^a$ representing
now $4 \times 4$ matrices. When $M,m>0$ or $M,m < 0$, these equations have now 
{\it two} solutions $q^{(1/2)}$ and $q^{(3/2)}$
corresponding (for positive masses) 
to the isospin projections $1/2$ and $3/2$. The projection $1/2$ 
gives a single vacuum state by the same token as the fundamental matter multiplet does. 
But the constant solution with $I_3 = 3/2$ can be transformed with the matrix
 \be
\label{Omega32} 
 \Omega_1^{3/2} = \exp \left\{ \frac {4 \pi i x}{3L} T^3 \right\}
 \ee
 such that periodicity of the matter fields is kept. On the other hand, 
neither $SU(2)$ nor $SO(3)$ matrices corresponding to \p{Omega32} are  periodic, 
and they need not to be: the only requirement is for
the configuration $\tilde{q}^{3/2} (x) = \Omega_1^{(3/2)}(x) \, q^{(3/2)}$ (supplemented by a certain constant gauge field 
$A_1^3$ ) to satisfy the periodic
boundary conditions and have zero classical energy.  
We obtain thus  {\it nine} classical states
\footnote{The quantum states with definite electric fluxes represent their linear combinations.} 
 \be
\label{ninestates}
 |0\rangle^{3/2}_{pq} \
 =\ \left(\Omega_1^{3/2}\right)^p  \left(\Omega_2^{3/2} \right)^q |0\rangle^{3/2} , \ \ \ \ \ \ \ \ p,q = 0,1,2 \,  
 \ee
where $|0\rangle^{3/2}$ is the classical vacuum with constant fields.  
Adding to this the state $|0\rangle^{1/2}$, we obtain altogether ten states, which coincides with $T_2(3/2)$.

Let now the theory involve several ${\cal N} = 2$ multiplets with different isospins $I_f$. Suppose that the
Lagrangian represents the pure ${\cal N} = 2$ SYMCS Lagrangian where the terms describing the interaction  between 
the gauge ${\cal N} = 2$ multiplet and the matter  ${\cal N} = 2$ multiplets endowed each with a real mass 
are added. Then the matter-induced shift of the index  is a sum of the shifts due to
 individual multiplets,
 \be
\label{mnogo}
I = |k| - 1 + \frac 12 \sum_f T_2(I_f) \, .
 \ee 

For rich enough matter content, one can write in the Lagrangian also  cubic ${\cal N} = 2$ invariant superpotentials. 
This can bring about extra Higgs vacuum states on the Higgs branches by the same mechanism as it does in 4 dimensions.

\subsection{${\cal N}=2$: Abelian theories.}

We will discuss here only the vectorlike Abelian theories. Chiral
theories can also be considered, but they involve certain complications 
\cite{ISnew}, which we do not want to come to grips with here.  
 The simplest theory of this kind involves the gauge ${\cal N} = 2$ multiplet 
$\{\Gamma_\alpha, \Sigma\}$ and a pair of matter multiplets $Q_f$ 
of the same mass and opposite charges. 
We write the Lagrangian in the full analogy 
with \p{fundN2},
\be
\label{LU(1)}
{\cal L} = \ \int d^2\theta  \left[
  \frac 1{2e^2} ( W_\alpha W^\alpha - 
D_\alpha  \Sigma \, D^\alpha \Sigma )  + \ \frac {i\kappa}2 
\left( \frac 12 W_\alpha \Gamma^\alpha -   \Sigma^2 \right) \right] 
\nonumber \\ 
+ \int d^2\theta  \left\{ \frac 1{e^2} \left[ -\frac 12    
\bar Q_f \, \nabla^\alpha
\nabla_\alpha Q_f + i   \Sigma (  \bar Q_1 Q_1  - \bar Q_2 Q_2 )   \right] -
i\xi \bar Q_f Q_f \right \}
  \, 
  \ee  
with $\nabla_\alpha Q_1 = (D_\alpha - \Gamma_\alpha) Q_1$ and 
$\nabla_\alpha Q_2 = (D_\alpha + \Gamma_\alpha) Q_2$.

The constant $\kappa$ is also quantized here but, in contrast to the 
non-Abelian case where it is already quantized at the level 
of pure SYMCS theory, the first line in \p{LU(1)} describes a free theory 
where $\kappa$ can be arbitrary. It has to be quantized only 
if the interactions with the matter are taken into account. Basically, 
the quantization  follows from the
requirement that the wave function stays invariant up to  
phase factors $e^{i\theta_1}, \, e^{i\theta_2}$ under the  transformations
 \be
\label{Abgauge}
G_1: \ 
A_1 (\vecg{x}) &\to& A_1  (\vecg{x})  + \frac {2\pi }L\, , \ \ \ \ \ \ \ 
Q_{1,2}(\vecg{x}) \to e^{\pm 2\pi i  x/L} Q_{1,2}(\vecg{x}) \nn
G_2: \ 
A_2 (\vecg{x}) &\to& A_2 (\vecg{x})  + \frac {2\pi }L\, , \ \ \ \ \ \ \ 
Q_{1,2}(\vecg{x}) \to e^{\pm 2\pi i  y/L} Q_{1,2}(\vecg{x}) \, 
   \ee 
 that respect
 the periodicity of $Q_f(\vecg{x})$ in a finite box. The transformations \p{Abgauge} look like  
gauge transformations, but [in contrast to \p{shift}] 
they are not contractible. 
Different phases $\{\theta_1, \theta_2\}$ correspond to different sectors in the Hilbert space that do not talk to each other.  
In each such sector, the  zero Fourier mode $A_j^{\bf (0)}$ lives effectively on the dual torus of size 
$\frac {2\pi}L$. To keep the spectrum of the Hamiltonian supersymmetric \cite{flux}, the magnetic flux on this torus,  
 \be 
\label{kAb}
\frac {\Phi}{2\pi} \ \equiv k \ = 2\pi \kappa  
 \ee
must be integer [note the difference in normalization compared to \p{IndpureN1} ]. 

The  Witten index of the pure ${\cal N} = 2$ supersymmetric Maxwell--Chern--Simons theory [the first line in \p{LU(1)}] put on 
the dual torus of size $2\pi/L$ is equal
to $|k|$. The matter fermions bring about the renormalization 
\footnote{For a chiral theory, on top of renormalizing the level $k$, also the effective Fayet-Illiopoulos term 
$\propto \int \Sigma \, d^2\theta$ can be generated. This is a complication we told about in the beginning of this subsection.}, 
 $k \to k - \left(\frac 12 + \frac 12 \right) {\rm sgn}(\xi) = 
k - {\rm sgn}(\xi)$. 
When $k\xi $ is positive, two extra  Higgs vacuum states (one for each matter flavor) should be added. This gives
 \be
\label{IndU(1)}    
I = |k| + 1 \, .
  \ee

Consider now the theory involving  an even number $2N_f$ of 
matter multiplets $Q_f, \bar Q_f$ that form  $N_f$ chirally symmetric pairs. We will assume that each such pair
has the same mass, $m_1 = m_2, \ldots , m_{2N_f-1} = m_{2N_f}$ and opposite  
 integer charges,  $Z_1 = -Z_2, \ldots, Z_{2N_f -1} = -Z_{2N_f}$,
  including also the unit charge 
\footnote{It is enough actually to require that the greatest common divisor of $\{Z_f\}$ is 1.} 
and the Lagrangian for each such pair
has the same form as in \p{LU(1)}. Then the index represents a sum \cite{ISnew}
  \be
\label{IndU(1)mnogo}    
I = |k| +  \frac 12 \sum_{f=1}^{2N_f} Z_f^2 \, .
  \ee
Indeed, the shift of $k$ due to the loop of the fermions carrying the charge $Z_f$ involves the factor $Z_f^2$. Also the number
of the Higgs states associated with the multiplet $f$ is equal to $Z_f^2$ : besides the Higgs vacuum $|0\rangle$ 
with constant fields $q_f$, there
exist also the vacua  
$$|0\rangle_{pq} =  [\Omega_1(x)]^{Z_f p}  [\Omega_2(y)]^{Z_f q} |0\rangle\, ,  \ \ \ \ \ \ \ \ \ \ 
p,q = 0, 1,\ldots, Z_f-1$$
 with  
 \be
\label{AbOmega}
 \Omega_1 \ =\ \exp\{2\pi i x/(Z_f L)\}, \ \ \ \ \ \ \ \ \ \ \ \Omega_2 \ =\ \exp\{2\pi i y/(Z_f L)\} \, .
 \ee
 $\Omega_{1,2}$ can be interpreted as ``fractional gauge transformation'' factors that would multiply the field
$q_g(\vecg{x})$ of unit charge.

\subsection{ ${\cal N}=2$: higher unitary groups.}

Besides the results \p{IndU(1)mnogo} and \p{mnogo} derived in \cite{ISnew} for 
the Abelian and $SU(2)$ theories, Intriligator and Seiberg also 
 conjectured the value of the index for higher unitary groups [Eq.(1.5) in their paper]. 
This represents a natural generalization of \p{mnogo} and respects certain claimed dualities 
\cite{Ken}.   We will derive it here by our method. 
The latter is the same as for $SU(2)$ and for $U(1)$ : performing all the necessary renormalisation 
and, when proper, taking into account 
Higgs vacuum states, we reduce 
the problem to evaluating the index in a pure ${\cal N} = 1$ SYMCS theory where the answer is known.

Consider  the simplest nontrivial example when the Lagrangian involves the gauge ${\cal N} = 2$ multiplet
and the fundamental matter multiplet and let first the gauge group be $SU(3)$.
The index of the pure  ${\cal N} = 1$ SYMCS theory with $SU(3)$ gauge group is known to be
 \be
\label{IndN1SU3}
I^{\rm SYMCS}_{{\cal N} = 1\, , \ SU(3)} \ =\ \frac 12 \left( k^2 - \frac 14 \right) \, .
  \ee
$k$ must be half-integer here. Supersymmetry is broken when $|k| = 1/2$.

 The ${\cal N} = 2$ SYMCS theory involves an extra adjoint matter multiplet $\Sigma$. Its mass is positive for $k > 0$ and negative for 
$k < 0$. The level is renormalized according to $k \to k - \frac 32 {\rm sgn}(k)$. Substituting this in \p{IndN1SU3}, one obtains
  \be
\label{IndN2SU3}
I^{\rm SYMCS}_{{\cal N} = 2\, , \ SU(3)} \ =\ \frac 12 ( |k| -2)(|k| -1)  \, .
  \ee
Supersymmetry is thus broken at $|k| = 1,2$. 

With the extra matter fundamental multiplet of positive mass, $k$ is renormalized as
   \be
\label{krenSU3matt}
k &\to& k - \frac 32 - \frac 12 = k-2 \, , \ \ \ \ \ \ \ \ \ \ \ \ \ \ \ \ \ k>0 \, , \nonumber \\
k &\to& k + \frac 32 - \frac 12  = k+1 \, , \ \ \ \ \ \ \ \ \ \ \ \ \ \ \ \ \ k<0 \, .
\ee
Substituting this in \p{IndN1SU3}, we obtain the following contributions to the index,
 \be
\label{IndSU3otren}
I &=& \frac 12 \left( k - \frac 52 \right) \left( k - \frac 32 \right)\, , \ \ \ \ \ \ \ \ \ \ \ \ \ k>0 \, , \nn
 I &=& \frac 12 \left( k + \frac 12 \right) \left( k + \frac 32 \right)\, , \ \ \ \ \ \ \ \ \ \ \ \ \ k<0 \, .
   \ee

To this, we must add the  vacua associated with nonzero Higgs vacuum expectation values. The {\it classical} vacua
 are the solutions to the same equation as \p{uslovmin}, but $t^a$ are now the $SU(3)$
generators. By a gauge rotation, one can bring $\sigma^a$ onto the maximal torus, $\sigma^a t^a = \sigma^3 t^3 + \sigma^8 t^8$. The second
equation in \p{uslovmin} then implies, in particular, that $\bar q t^{1,2,4,5,6,7} q = 0$, and this is possible if only one component among 
$q_1, q_2, q_3$ is nonzero. Let the nonvanishing component be $q_1$ and let it be real. 
Choose $m$ to be positive. Let $M$ be also positive. After a simple algebra, we obtain a solution 
 \be
\label{HiggsSU3}
 q_1 = \sqrt{3mM}, \ \ \ \ \ \ \ \ \ \ \ \ \sigma^a t^a = m \, {\rm diag}\left( 1, - \frac 12, -\frac 12 \right) \, .  
 \ee

If assuming that $q_1 = q_3 = 0$, we would obtain the solution
 $$
 q_2 = \sqrt{3mM}, \ \ \ \ \ \ \ \ \ \ \ \ \sigma^a t^a = m \, {\rm diag}\left(- \frac 12 , 1, -\frac 12 \right) \, 
 $$
which represents a gauge copy of \p{HiggsSU3}. The same for the case $q_1=q_2 = 0$.
There is {\it no} solution when $M < 0$. Thus, for the Higgs fields in fundamental representation, we obtain only {\it one} solution.
\footnote{For higher representations, one would obtain a nontrivial multiplicity of classical 
Higgs vacua. For example, for the matter 
in the adjoint representation (the theory discussed in the Appendix), 
the multiplicity is 9.}

It would be wrong, however, to add just 1 
to the first line in \p{IndSU3otren}. For $SU(2)$ and for $U(1)$, 
Higgs v.e.v.'s broke gauge group completely, there were no light fields left, 
and the corresponding vacua were isolated. But in the case of $SU(3)$, 
the v.e.v. \p{HiggsSU3} leaves the $SU(2)$ subgroup of $SU(3)$ unbroken. 
The BO dynamics is nontrivial in this case. It
corresponds to the pure ${\cal N} = 1$ SYMCS theory with the $SU(2)$ gauge 
group and the renormalized coupling 
$k \to k - \frac 32$ [see Eq.\p{krenSU2}]. The index of this theory is equal to
$k-\frac 32$. Adding {\it this} to the first line in \p{IndSU3otren}, we
obtain the final universal result     
  \be
\label{IndSU3matt} 
 I^{SU(3)} \ =\ \frac 12 \left( |k| - \frac 32 \right)
 \left( |k| - \frac 12 \right)\, .
 \ee
The same follows from  (1.5) of Ref.\cite{ISnew}.
 The supersymmetry is broken when $|k| = 1/2$ or $|k| = 3/2$.

Consider now a generic $SU(N)$ group. In this case, the index of the pure $N=1$ SYMCS theory is 
     \be
  \label{IndN1SUN}  
   I^{{\rm pure}\ {\cal N}= 1}_{SU(N)} \ =\ \frac 1{(N-1)!} \prod_{j = -\frac N2 +1}^{\frac N2 -1} (k-j) \, . 
    \ee

The effective BO theory associated with zero classical Higgs v.e.v.'s is the theory with renormalized $k$: 
   \be
\label{krenSUNmatt}
k &\to& k - \frac {N+1} 2  \, , \ \ \ \ \ \ \ \ \ \ \ \ \ \ \ \ \ k>0 \, , \nonumber \\
k &\to& k + \frac {N-1} 2  \, , \ \ \ \ \ \ \ \ \ \ \ \ \ \ \ \ \ k<0 \, .
\ee   
 The corresponding contributions to the index are
 \be
\label{IndSUNotren}
 I &=&  \frac 1{(N-1)!} \prod_{j = -\frac N2 +1}^{\frac N2 -1} \left( k -  \frac {N+1} 2 -j \right) \, .
\ \ \ \ \ \ \ \ \ \ \ \ \ \ \ \ \ \ \ \ k>0 \,, \nn
I &=&  (-1)^{N-1} \frac 1{(N-1)!} \prod_{j = -\frac N2 +1}^{\frac N2 -1} \left( k +  \frac {N-1} 2 -j \right) \, .
\ \ \ \ \ \ \ \ \ \ \ \ \ \ \ \ \ \ \ \ k<0 \, . 
   \ee

The classical Higgs vacuum represents, again, a solution of \p{uslovmin}. As was also the case for $N = 2,3$,
 a  unique up to a gauge transformation
 solution exists for positive, 
but not for negative $k$. It can be written as
\footnote{A mathematician would recognize in 
$\sigma^a t^a$ in \p{HiggsSUN}  a {\it fundamental coweight} - an element of the Cartan subalgebra orthogonal
to all simple coroots but one.}
  \be
\label{HiggsSUN}
 q_1 = \sqrt{\frac {2NmM}{N-1}}, \ \ \ \ \ \ \ \ \ \ \ \ \sigma^a t^a = m \, {\rm diag}\left( 1, - \frac 1{(N-1)}, \cdots ,
 -\frac 1{(N-1)} 
\right) \, .  
 \ee  
 These v.e.v's break the group $SU(N)$ down to $SU(N-1)$. The contribution to the index associated with
the classical vacuum \p{HiggsSUN} is given again by \p{IndN1SUN} with $N$ replaced by $N-1$ and $k$ by 
$k - \frac N2$, 
  \be
\label{IndHiggsSUN}
I^{SU(N)}_{\rm Higgs} &=& \frac 1{(N-2)!} \prod_{j = -\frac {N-1}2 +1}^{\frac {N-1} 2 -1} \left( k - \frac N 2 -j \right)
 \, .
   \ee

Adding this (for $k > 0$) to \p{IndSUNotren}, we obtain the universal result
  \be
\label{IndSUNfinal}
 I^{SU(N)} &=&  \frac 1{(N-1)!} \prod_{j = -\frac N2 +1}^{\frac N2 -1} \left( |k| + \frac 12 - \frac {N} 2 -j \right) \, .
   \ee

 Note that, for all ${\cal N} = 2$ theories considered so far, 
the index is the same for positive
and negative $k$ and for the positive and negative masses whereas {\it a priori} one could expect only
the  symmetry with respect to the
spatial parity transformation that changes the signs of $k$ and of all masses simultaneously. An interesting 
explanation for the symmetry with respect
to  mass sign flip with given $k$ (and hence with respect to the sign flip of $k$ with given $m$) 
was suggested in \cite{ISnew}. 
Basically, they argued that one can add to the mass the size of one of the cycles 
of the dual torus multiplied by $i$ 
to obtain  a complex holomorphic parameter on which the index of an ${\cal N} = 2$ theory 
should not depend. And hence it should not depend on the real part of this
parameter (the mass). To my mind, it is still dangerous to pass the point $m=0$ where the 
index is not defined and this argument thus 
lacks rigour. But, at least for the unitary groups with fundamental matter 
considered above and for the $SU(3)$ theory with
adjoint matter considered in the Appendix, 
the symmetry with respect to mass sign flip is there, indeed. 
\footnote{We emphasize that this is all 
an  ${\cal N} = 2$  specifics. For ${\cal N} = 1$ theories, 
there is no such symmetry, 
 see e.g. Fig.2.}
It would be interesting to construct a rigourous proof of this fact.   

Accepting the existence of this symmetry, it is not difficult to derive a 
generalization of \p{mnogo}
for a $SU(N)$ theory   
 involving several matter multiplets in the representations $R_f$ with real 
masses and without extra Yukawa couplings. 
It is given by the same formula  \p{IndSUNfinal} where we should
replace 
 \be
\label{replacek}
 |k| + \frac 12 \ \to \ |k| + \frac 12 \sum_f T_2(R_f) \, .
 \ee
For the negative $k$ where no Higgs states contribute, the R.H.S. of \p{replacek} is just the net renormalization 
of $k$ due to the matter multiplets, while for positive $k$ 
the result can be restored using the symmetry mentioned above.

The expression \p{IndSUNfinal} with the substitution \p{replacek} 
coincides with (1.5) in \cite{ISnew}, as announced.

This analysis can be extended to 
an arbitrary gauge group where the index for the pure SYMCS ${\cal N} =1$ theory is known. 
Besides
unitary groups, the explicit expressions  were derived for the symplectic groups and for $G_2$. 
For the group $Sp(2r)$ of rang $r$ and for positive $k$, the index is 
 \be
\label{ISp2r}
 I^{\rm SYMCS}_{{\cal N} = 1} [Sp(2r)] \ =\ \left( \begin{array}{c} k + \frac {r-1}2 \\ r \end{array} \right) \, .
 \ee 
For the negative $k$, the index is restored via $I(k) = (-1)^r I(-k)$. 

The index for $G_2$ is
\be
\label{IG2}
 I^{\rm SYMCS}_{{\cal N} = 1} [G_2] \ =\ \left\{  \begin{array}{c} \frac {k^2}4 \ \ \ \ \ \ \ \ \ {\rm for\ even\ } k \\  
\frac {k^2-1}4 \ \ \ \ \ \ \ \ \ {\rm for\ odd\ } k \end{array} \right. \ .
 \ee 
The results \p{ISp2r} and \p{IG2} are obtained from the tree-level expressions (1.6) and (1.7) of Ref.\cite{ja1} 
with taking into account the renormalization $k \to k - c_V/2$ due exclusively to fermion loops.
For the ${\cal N} = 2$ theories in interest, the result is obtained by
taking into account, for negative $k$,  its further renormalization due to matter fermion fields and 
assuming that the result
for  positive $k$ is the same.

\section{Acknowledgements.}
I am indebted to Z. Komargodski and N. Seiberg for useful discussions and to 
K. Intriligator for illuminating  correspondence and many 
valuable comments.

\section*{Appendix: $SU(3)$ with adjoint matter.}
\setcounter{equation}0
\def\theequation{A.\arabic{equation}}

We present here the accurate calculation of the index in the $SU(3)$ ${\cal N}=2$ theory with an extra 
adjoint matter multiplet. A general formula
described at the end of the paper gives in this case,
  \be
\label{IndSU3adj}
 I^{SU(3)}_{\rm adj. \ matt.} = \ \frac 12 (|k|+1)(|k| +2) \, .  
   \ee
(cf. the $SU(2)$ expression in \p{IndN2adj2} ).

For negative $k$ (and positive mass), the derivation is easy. We have just to substitute in \p{IndN1SU3} the renormalized $k$,
 $$k \to k +\left. \frac 32 \right|_\Sigma - \left. 3 \right|_{\rm matter}   = k - \frac 32 \, .$$
For positive $k$, an analogous procedure gives a contribution
 \be
\label{k-45}
 I \ =\ \frac 12 (k-4)(k-5) \, . 
 \ee
One should add to this Higgs states. To count them, we have to solve an $SU(3)$ generalization of \p{uslovminadj} 
which is convenient to present in the matrix form,
  \be
\label{uslovmatr}
 m\phi &=& [\phi, \sigma], \nonumber \\ 
 m\phi^\dagger &=& -[\phi^\dagger, \sigma]\, , \nonumber \\
M\sigma &=& [\phi^\dagger, \phi] \, .   
  \ee
It is clear that a solution of \p{uslovmatr} describes an embedding 
$su(2) \subset su(3)$. There are two such distinct embedding:
a natural embedding, like
\be
 \label{s-emb}
 \sigma \propto t^3, \ \ \ \ \phi \propto t^{1+i2}, \ \ \ \ \ \ \phi^\dagger \propto t^{1-i2} \, , 
 \ee
 which leaves the $U(1)$ group associated with the generator
$t^8$ unbroken, and also the nonstandard (but very well known, of course) 
embedding which breaks the gauge group completely  
(an embedding with the trivial centralizer, as a mathematician would say). 

By a gauge rotation, the latter can be brought to
the form
 \be
\label{ns-emb}
\sigma  \propto  \left( \begin{array}{ccc} 1 & 0 & 0 \\ 0 & 0 &0 \\ 0 & 0 & -1 \end{array} \right)\, , \ \ \ 
\phi   \propto \left( \begin{array}{ccc} 0 & 1 & 0 \\ 0 & 0 &1 \\ 0 & 0 & 0 \end{array} \right)\, , \ \ \
\phi^\dagger   \propto  \left( \begin{array}{ccc} 0 & 0 & 0 \\ 1 & 0 & 0 \\ 0 & 1 & 0 \end{array} \right)\,  .
 \ee
As the gauge group is broken completely, the vacuum \p{ns-emb} is isolated. Besides  \p{ns-emb}, 
there are $3^2 -1 = 8$ other isolated Higgs vacua obtained by  non-contractible quasi-gauge transformations that
keep periodicity of all adjoint fields, like in \p{largegauge}. The net contribution to the index is thus 9.

Let us discuss now the contribution to the index due to the standard embedding \p{s-emb}. As the $U(1)$ group is left unbroken,
we have to count the index in the corresponding effective Abelian theory. Its nontrivial part 
 involves the field $A_\mu^8(x) \to A_\mu(x)$, 
two pairs of the massive charged fields  $\phi_{1,2}$ and $\tilde{\phi}_{1,2}$ 
that come from the 
components $\phi^{4,5}$ and $\phi^{6,7}$ of the full theory (there are also neutral fields that decouple)
 and fermion superpartners. 
The bosonic part of the effective Lagrangian is   
  \be
\label{LeffAb}
{\cal L} \ =\ -\frac 1{4g^2} F_{\mu\nu}^2  +  \frac 1{2g^2} \left| 
\left( \partial_\mu - \frac {i\sqrt{3}}2 A_\mu \right)  \phi_1 \right|^2 + 
\frac 1{2g^2} \left| \left( \partial_\mu + \frac {i\sqrt{3}}2 A_\mu \right)  \phi_2 \right|^2 \nonumber \\ 
+ 
\frac 1{2g^2} \left| \left( \partial_\mu - \frac {i\sqrt{3}}2 A_\mu \right)   \tilde{\phi}_1 \right|^2 
+ \frac 1{2g^2} \left| \left( \partial_\mu + \frac {i\sqrt{3}}2 A_\mu \right)  \tilde{\phi}_2 \right|^2 + 
\frac \kappa 2 \epsilon^{\mu\nu\rho} A_\mu \partial_\nu A_\rho \, .
  \ee
 The coefficient $\frac {\sqrt{3}}2$ comes from the commutators $[t^8, t^{4+i5}] = \frac {\sqrt{3}} 2 t^{4+i5},
\ [t^8, t^{6+i7}] = \frac {\sqrt{3}} 2 t^{6+i7} $.
It is convenient to rescale $A_\mu \to B_\mu = A_\mu \frac {\sqrt{3}} 2$ such that, when also rescaling  $g^2$ 
and $\phi$ in a proper way, 
 the effective theory is brought to the form discussed in the first part of Sect. 4.7,  
involving two pairs of charged fields of opposite unit charges.

The rescaling $A_\mu \to B_\mu$ modifies the Chern-Simons coefficient, which is now $\kappa_{\rm eff} = \frac {4\kappa}3$.
The effective theory \p{LeffAb} describes the dynamics in the vicinity of the Higgs minimum \p{s-emb}. 
It makes sense to consider it only if the deviations from the minimum (the fields $\phi_{1,2}$ and $\tilde{\phi}_{1,2}$ )
are small. Thus, we have only to count the contribution to the index due to the region near the origin and disregard Higgs vacua 
that are also present in \p{LeffAb}. This contribution is 
 \be
\label{Ikeff}
I = k_{\rm eff} - 2 
 \ee 
 with
\be 
\label{keff}
k_{\rm eff} \ =\ 2\pi \kappa_{\rm eff} =\ \frac {8\pi \kappa}3  \ =\ \frac {2k}3 \, ,
 \ee
where $k = 4\pi \kappa$ is the level in the {\it original} theory.

The contribution \p{Ikeff} should be further multiplied by 9 (the vacuum \p{s-emb} has 8 twisted copies). 
This gives 
 \be
\label{Istand}
I_{\rm stand. \ embedding} \ =\ 6(k-3) \, .
 \ee 
  Note that the contribution of an {\it individual} state \p{s-emb}  is integer only if $k = 3l$. 
For $k = 3l+1,
3l+2$, the contribution \p{Ikeff} is fractional and does not have as such a  lot of meaning, only the full 
contribution 
\p{Istand} does. The situation is similar here to what we encountered discussing loop corrections for the pure ${\cal N} = 1$ SYMCS theory. 
The induced fluxes 
in each corner \p{corners} are half-integer and it makes no sense to talk about a contribution to the index coming from
an individual corner. The full quantum wave functions \p{koren} know about all four of them. 
   
Adding to \p{Istand} nine states 
associated with the vacuum \p{ns-emb}, we obtain all together the contribution $6k-9$ coming from the 
Higgs vacua. And this together with \p{k-45} leads to \p{IndSU3adj}, as anticipated.


\begin{thebibliography}{96}

\bibitem{ISnew}  K. Intriligator and N. Seiberg, 
{\it Aspects of 3d ${\cal N} = 4$ Chern-Simons-matter theories}, JHEP {\bf 1307}, 079 (2013),
  arXiv:1305.1633 [hep-th].

\bibitem{Wit99} E. Witten, {\it Supersymmetric index of three-dimensional gauge theory} in: 
   [Shifman, M.A. (ed.) The many faces of the superworld, p.156], arXiv: hep-th/9903005.

\bibitem{N>1} K. Ohta, {\it Supersymmetric index and $s$ rule for type IIB branes }, 
JHEP {\bf 9910}, 006 (1999) arXiv:hep-th/9908120;
   O. Bergman, A. Hanany, A. Karch, and B. Kol, {\it Branes and supersymmetry
 breaking in three-dimensional gauge theories}, JHEP {\bf 9910}, 036 (1999)
   arXiv:hep-th/9908075.

\bibitem{ja1}  A.V. Smilga, {\it Witten index in supersymmetric 3d theories revisited}, 
JHEP {\bf 1001} (2010) 086,   arXiv:0910.0803 [hep-th].

\bibitem{ja2}  A.V. Smilga, {\it Once more on the Witten index of 3d supersymmetric YM-CS theory}, 
JHEP {\bf 1205} (2012) 103,  arXiv:1202.6566 [hep-th].

\bibitem{ABJM} J. Bagger and N. Lambert, {\it Gauge symmetry 
and supersymmetry of multiple M2-branes}, Phys. Rev. {\bf D77}  (2008) 
065008, arXiv:0711.0955 [ hep-th]; 
O. Aharony, O. Bergman, D.L. Jafferis, and J. Maldacena, {\it ${\cal N} = 6$ superconformal 
Chern-Simons-matter theories, M2-branes and their gravity duals}, 
JHEP {\bf 0810}:091 (2008), arXiv:0806.1218 [hep-th]. 

\bibitem{Spir} G. Romelsberger, {\it Counting chiral primaries in ${\cal N} = 1$, 
$d=4$ superconformal field theories}, Nucl. Phys. {\bf B747} (2006) 329, arXiv:hep-th/0510060; 
V.P. Spiridonov and G.S. Vartanov,  {\it Superconformal indices 
for ${\cal N} =1$ theories with multiple duals}, 
Nucl. Phys. {\bf B824} (2010) 192,  arXiv:0811.1909 [hep-th].

\bibitem{Wit82} E. Witten, {\it Constraints on supersymmetry breaking}, Nucl. Phys. {\bf B202} (1982) 253.

\bibitem{Witort} E. Witten, {\it Toroidal compactification without vector structure}, 
JHEP {\bf 9802} (1998) 006, arXiv:hep-th/9712028.

\bibitem{Rosly} A. Keurentjes, A. Rosly, and A.V. Smilga, {\it Isolated vacua in supersymmetric Yang-Mills theories},
 Phys. Rev. {\bf D58} (1998) 081701, 
      hep-th/9805183.

\bibitem{jaKac} V.G. Kac and A.V. Smilga, {\it Vacuum structure in supersymmetric Yang-Mills theories 
with any gauge group} in:
     [Shifman, M.A. (ed.) The many faces of the superworld, p.185], arXiv:hep-th/9902029.

\bibitem{Keur} A. Keurentjes, {\it Nontrivial flat connections on the 3 torus I: $G(2)$ and the orthogonal groups},
JHEP {\bf 9905} (1999) 001, arXiv:hep-th/9901154; 
{\it Nontrivial flat connections on the 3 torus II: The exceptional groups F4 and E6,E7,E8},
JHEP {\bf 9905} (1999) 014, arXiv:hep-th/9902186. 

\bibitem{jachiral} A.V. Smilga, {\it Vacuum structure in the chiral supersymmetric quantum electrodynamics}, 
JETP {\bf 64} (1986) 8; 
        B.Yu. Blok and A.V. Smilga, {\it Effective zero mode Hamiltonian in supersymmetric 
chiral non-Abelian gauge theories},  Nucl. Phys. {\bf B287} (1987) 589.

\bibitem{ISold}  K. Intriligator and N. Seiberg, {\it Phases of ${\cal N} = 1$ supersymmetric gauge theories in four 
dimensions}, Nucl. Phys. {\bf B431} (1994) 551, arXiv:hep-th/9408155.

\bibitem{GVY} A. Gorsky, A. Vainshtein, and A. Yung, {\it Deconfinement at the Argyres-Douglas point
in $SU(2)$ gauge theory with broken supersymmetry}, Nucl. Phys. {\bf B584} (2000) 197, arXiv:hep-th/0004087.

\bibitem{ADS} I. Affleck, M. Dine, and N. Seiberg, {\it Dynamical supersymmetry breaking in chiral theories}, 
Phys. Lett. {\bf B137} (1984) 187. 

\bibitem{SW94} N. Seiberg and E. Witten, {\it Monopoles, duality and chiral symmetry breaking 
in ${\cal N} = 2$ supersymmetric QCD}, Nucl. Phys.  {\bf B431} (1994) 484, arXiv:hep-th/9408099.


\bibitem{G2} A. Smilga, {\it 6+1 vacua in supersymmetric QCD with $G_2$ gauge group}, 
Phys. Rev. {\bf D58} (1998) 105014, arXiv:hep-th/9801078.


\bibitem{Gates} S.J. Gates, M.T. Grisaru, M. Rocek, and W. Siegel, 
{\it Superspace or one thousand and one lessons in supersymmetry}, Front. Phys. {\bf 58} (1983) 1-548, arXiv:hep-th/0108200.

 \bibitem{realmass} H. Nishino and S.J. Gates, {\it Chern-Simons theories with supersymmetries in three-dimensions}, 
Int. J. Mod. Phys. {\bf A8} (1993) 3371; 
O. Aharony et al, {\it Aspects of ${\cal N} =2$ supersymmetric gauge theories in three-dimensions},  
Nucl. Phys. {\bf B499} (1997) 67, arXiv:hep-th/9703110;
 J. de Boer, K. Hori, and Y. Oz, {\it Dynamics of ${\cal N} =2$ supersymmetric gauge 
theories in three-dimensions}, Nucl. Phys. {\bf B500} (1997), arXiv:hep-th/9703100.

\bibitem{Ivanov} E.A. Ivanov, {\it Chern-Simons matter systems with manifest ${\cal N} =2$ supersymmetry}, 
Phys. Lett. {\bf B268} (1991) 203.

\bibitem{Henningson} M. Henningson, {\it Ground states of supersymmetric Yang-Mills-Chern-Simons theory}, 
JHEP {\bf 1211} (2012) 013, arXiv:1209.1798 [hep-th].

\bibitem{DJT} S. Deser, R. Jackiw, and S. Templeton, {\it Topologically massive gauge theories}, 
Ann. Phys. (NY) {\bf 140} (1982) 372.

\bibitem{Novikov} B.A. Dubrovin, I.M. Kri\v{c}ever, and S.P. Novikov, 
{\it The Schr\"odinger equation in a periodic field and 
Riemann surfaces},  Soviet Math Dokl. {\bf 229} (1976) 15;
B.A. Dubrovin and S.P. Novikov, {\it Ground states of a two-dimensional electron in periodic magnetic field}, 
Soviet Phys. JETP {\bf 79} (1980) 1006.

\bibitem{ja3} A.V. Smilga, {\it Vacuum structure in 3d supersymmetric gauge theories}, 
Usp. Fiz. Nauk {\bf 184} (2014) 163, Phys. Usp. {\bf 57} (2) (2014), 
 arXiv:1312.1804 [hep-th]. 

\bibitem{Pi} K. Gawedzki and A. Kupiainen, {\it Coset construction from functional integrals}, 
Nucl. Phys. {\bf B320} (1989) 625;
S. Elitzur, G. Moore, A. Schwimmer, and N. Seiberg, {\it Remarks on the canonical quantization 
of the Chern-Simons-Witten theory}, Nucl. Phys. {\bf B326} 
(1989) 108; J.M.F. Labastida and A.V. Ramallo, {\it Operator formalism for Chern-Simons theories}, 
Phys. Lett. {\bf B227} (1989) 92.

\bibitem{Kao2Lee} H.-C. Kao, Kimyeong Lee, and Taelin Lee, 
{\it The Chern-Simons coefficient in supersymmetric Yang-Mills Chern-Simons theories}, 
Phys. Lett. {\bf B373} (1996) 94, arXiv:hep-th/9506170.

\bibitem{Hooft} G. 't Hooft, {\it A property of electric and magnetic flux in non-Abelian gauge theories}, 
Nucl. Phys. {\bf B153} (1979) 141.

\bibitem{flux} A.V. Smilga, {\it Non-integer flux --- why it does not work}, 
J. Math. Phys. {\bf 53} (2012) 042103,   arXiv:1104.3986 [math-ph].

\bibitem{Ken} K. Intriligator, {\it private communication}. 


\end{thebibliography}
\end{document}